\documentclass{emulateapj}
\usepackage{graphicx}
\usepackage{amssymb}
\usepackage{amsmath}
\shorttitle{Gravitational Infall onto Filaments}
\shortauthors{Heitsch}

\newcommand{\f}{\frac}
\begin{document}
\title{Gravitational Infall onto Molecular Filaments}

\author{Fabian Heitsch\altaffilmark{1}}
\altaffiltext{1}{Department of Physics and Astronomy, University of North Carolina Chapel Hill, Chapel Hill, NC 27599-3255}
\email{fheitsch@unc.edu}

\begin{abstract}
Two aspects of filamentary molecular cloud evolution are addressed: (1) Exploring analytically the role of the
environment for the evolution of  filaments demonstrates that considering them in isolation (i.e. just
addressing the fragmentation stability) will result in unphysical conclusions about the filament's properties.
Accretion can also explain the observed decorrelation between FWHM and peak column density.
(2) Free-fall accretion onto finite filaments can lead to the characteristic ``fans'' of infrared-dark
clouds around star-forming regions. The fans may form due to tidal forces mostly arising
at the ends of the filaments, consistent with numerical models and earlier analytical studies. 
\end{abstract}


\section{Motivation}
Recent observations have re-emphasized the prominence of ``filamentary'' (at minimum, non-spherical) structures
in molecular clouds \citep{2010A&A...518L.102A,2010A&A...518L.100M,2011A&A...529L...6A}. 
While certainly a confirmation of earlier findings 
\citep{1979ApJS...41...87S}, higher observational resolution probes successively deeper into
the structure, and thus into the physical parameter space occupied by these ``filaments''.

Filaments are the most beneficial geometry for local gravitational collapse to occur before
global collapse \citep{2011ApJ...740...88P}. The global free-fall time increases with respect to the (spherical) three-dimensional
case \citep{2012ApJ...744..190T}, and -- equivalently -- the ratio of the local over the global collapse timescale decreases with 
reduced dimension \citep{2011ApJ...740...88P}. 

Self-gravitating molecular clouds rarely exist in complete isolation. \citet{2009ApJ...704.1735H} showed for model clouds
of similar parameters as the Pipe Nebula \citep{2008ApJ...672..410L} that once the clouds are not considered in isolation, the
gravitational potential generated by the cloud mass will lead to accretion. In fact, Pipe-like clouds are likely to 
accrete their own mass within a dynamical (read ``free-fall'') time. \citet{2009ApJ...704.1735H} also demonstrate that while
their models show clear infall and accretion, the kinematic signatures of such motions can be easily hidden if
the medium is structured, or by projection effects. Observed morphologies and kinematics suggest gas infall,
from the filament-core level \citep[][see also e.g. \citealp{2011A&A...533A..34H}]{2013ApJ...766..115K}, on to low-mass clouds 
\citep[][for Taurus]{1999ApJ...527..285B,2013A&A...550A..38P} up to clouds harbouring massive star-formation regions 
\citep{2005IAUS..227..151M,2010A&A...520A..49S,2012A&A...543L...3H,2013ApJ...766...68P}. These studies also suggest that the prominent filaments
are hosting most of the star formation, and thus geometric properties of filaments could control the star formation process
to a large extent. \citet{2011A&A...529L...6A} show that for their sample
of filaments, the filament width does not depend on the central column density, and speculate that this could be an effect
due to accretion. Gas accretion {\em along} filaments has also been
predicted based on simulations \citep{2001MNRAS.327..715B}.

The notion that molecular clouds do not exist in isolation, and thus will feed on their environment, raises two issues
that we explore for a better understanding of molecular cloud evolution.
\paragraph{(1) The Fate of an Accreting Filament} Gas accretion onto a filament will not only grow the cloud's mass
and thus reduce the fragmentation timescale, but it also will drive internal turbulence \citep{2010A&A...520A..17K}.
The relevance of accretion over fragmentation will be explored in \S\ref{s:accretion}. Already a comparison of the characteristic
timescales shows that accretion ought not to be neglected. 
\paragraph{(2) Free-fall Accretion onto a Filament} Accretion onto a filament will lead to characteristic kinematic
signatures that will be explored in \S\ref{s:tidal}. Specifically, tidal forces around a massive filament can cause 
``gravitational streamers'', i.e. fan-like structures converging at the ends of a filament.

The considerations presented here are purely analytical, and they are somewhat speculative. The goal is to explore an 
extreme view, namely that free-fall accretion governs the evolution of molecular clouds.
\S\ref{s:discussion} highlights some of the arguments for and against this assumption. 

\section{Evolution of an accreting filament}\label{s:accretion}

In this model, the evolution of an accreting filament is 
determined by two timescales: the gravitational fragmentation 
timescale along the axis of the filament, and the accretion timescale, 
i.e. the timescale it takes the filament to reach line masses 
\begin{equation}
  m>m_{cr}\equiv \frac{2c_s^2}{G}=16.3\left(\f{T}{10\mbox{K}}\right)\,\mbox{M}_\odot\mbox{ pc}^{-1}
  \label{e:mlinecrit}
\end{equation}
above the critical line mass $m_{cr}$ \citep[e.g.][]{1964ApJ...140.1056O}, or -- in the nomenclature
of \citet{2012A&A...542A..77F} -- to reach
\begin{equation}
  f\equiv \frac{m}{m_{cr}} = \frac{mG}{2c_s^2} > 1.
  \label{e:fcrit}
\end{equation}
We will denote the isothermal sound speed by $c_s$, and the line mass by $m$.

\subsection{Comparison of Instantaneous Timescales}
The most unstable scale of an isothermal, hydrostatic cylinder of infinite length is given by
\begin{eqnarray}
  \lambda_{max}&=&\f{20 c_s}{(4\pi G\rho_c)^{1/2}}\nonumber\\
              &=&0.67\left(\f{T}{10\mbox{ K}}\right)^{1/2} \left(\f{n_c}{10^4\mbox{ cm}^{-3}}\right)^{-1/2}\mbox{ pc}\label{e:lmax},
\end{eqnarray}
with the central density $\rho_c=\mu m_H n_c$.
The cylinder will fragment at this scale on a timescale of
\begin{eqnarray}
  \tau_f&=&\f{3}{\sqrt{4\pi\,G\,\rho_c}}\label{e:tau_f}\\
        &=& 5.24\times 10^5\,\left(\f{n_c}{10^4\mbox{ cm}^{-3}}\right)^{-1/2}\,\mbox{yrs},\nonumber
\end{eqnarray}
\citep{1987PThPh..77..635N,1995ApJ...438..226T}, due to a varicose or sausage
instability \citep[see][for an application to the ``Nessie'' filament; also \S\ref{sss:nessie}]{2010ApJ...719L.185J}.

The free-fall accretion timescale onto a filament can be derived 
by the ratio of the line mass $m$ over the accretion rate 
$\dot{m}$ through a cylinder circumference $2\pi R$,
\begin{equation}
  \tau_a = \f{m}{\dot{m}} = \f{m}{2\pi R\rho v_R}.
\end{equation}
Following the discussion of \citet[][see also Palmeirim et al. 2013]{2009ApJ...704.1735H} to 
determine the accretion velocity $v_R$, 
the gravitational acceleration of a gas parcel in free-fall onto a 
cylinder with line mass $m$ is given by 
\begin{equation}
  a_R = -\f{2Gm}{R}.
\end{equation}
Assuming a steady-state infall, this yields a radial velocity profile via
\begin{equation}
  v_R\f{d}{dR}v_R = -\f{2Gm}{R},
\end{equation}
namely
\begin{equation}
  v_R = 2\left(Gm\,\ln\f{R_{ref}}{R}\right)^{1/2}.
  \label{e:vrad}
\end{equation}
Four parameters remain to be determined: the parcel density
$\rho$, the line mass $m$, and the radii $R_{ref}$ and $R$. The 
ambient gas density $\rho_{ext}$ will serve as the parcel density, assuming
an average over a highly structured medium. The line mass $m$ is
set to the critical value for now -- awaiting a more detailed
treatment in the next section. The choice of the radii is trickier.
For now, suffice it to say that $v_R$ depends only weakly on $R_{ref}/R$, 
so that this choice is of lesser importance. The integration constant $R_{ref}$
can be identified with the starting position of the fluid parcel, located at --
somewhat arbitrarily -- $R_{ref}=2$~pc.
The radius $R$ itself is set
to a characteristic filament width from observations, e.g.
a few tenths of a parsec \citep[e.g.][]{2011A&A...529L...6A}.

Combining all expressions leads to 
\begin{eqnarray}
  \tau_a = 8.31\times 10^5& &\left(\f{T}{10\mbox{ K}}\right)^{1/2}
                          \left(\f{n_0}{100\mbox{cm}^{-3}}\right)^{-1}\nonumber\\
                 &\times& \left(\f{R}{\mbox{pc}}\right)^{-1}\left(\ln\f{R_{ref}}{R}\right)^{-1/2}
  \label{e:tau_a}
\end{eqnarray}

Thus, the accretion timescale (eq.~\ref{e:tau_a}) and the fragmentation timescale 
(eq.~\ref{e:tau_f}) are of the same order. Filaments that do not exist in 
isolation accrete gas on similar timescales under which 
they fragment. Yet, the above estimate suffers from the somewhat random choice of
parameters, begging a more detailed discussion.

\subsection{Time-dependent Timescales}

The next step beyond considering instantaneous timescales involves integrating
the expression for the mass accretion rate using equation~\ref{e:vrad},
\begin{eqnarray}
  \f{dm}{dt}&=&2\pi R\rho_{ext} v_R\nonumber\\
            &=&4\pi\rho_{ext} R\left(G m(t)\,\ln\left(\f{R_{ref}}{R}\right)\right)^{1/2}
  \label{e:dmdt}
\end{eqnarray}
Thus, the accretion rate
will depend on the evolutionary stage of the filament via the line mass $m(t)$, and
on the surface radius $R$. The latter is set to the filament radius. This
would be formally infinite for an isothermal, hydrostatic cylinder. Yet since 
filaments do not exist in vacuo, it seems reasonable to define the (outer) filament
radius $R_f$ as the radius $R$ at which the radial density profile
\begin{equation}
  \rho(R) = \rho_c\left(1+\left(\frac{R}{R_0}\right)^2\right)^{-p/2}
  \label{e:rhoR}
\end{equation}
has dropped to the ambient density $\rho_{ext}$, i.e.
\begin{equation}
  R_f = R_0\left(\left(\f{\rho_c}{\rho_{ext}}\right)^{2/p}-1\right)^{1/2}.
  \label{e:Rf}
\end{equation}
For an isothermal cylinder, $p=4$. The characteristic radius $R_0$ (equivalent to
$R_{flat}$ in \citet{2011A&A...529L...6A}) is set to the isothermal
core radius \citep{1964ApJ...140.1056O}, 
\begin{equation}
  R_0^2 = \frac{2c_s^2}{\pi G \rho_c} = \f{m_{cr}}{\pi \rho_c},
  \label{e:R0}
\end{equation}
using equation~\ref{e:mlinecrit}.
Note that this expression differs by a factor of $8$ from that given by 
\citet{1964ApJ...140.1056O}. 

Unlike in the analysis of pressurized, hydrostatic isothermal cylinders by \citet{2012A&A...542A..77F}, we specify the
steepness of the filament profile explicitly by setting $p$, instead of relying on an external pressure 
to flatten the profile. We discuss the rationale and physical interpration in \S\ref{sss:differences}.

The line mass $m(t)$ can be found by integrating over the density profile and by using equations~\ref{e:Rf} and \ref{e:R0},
\begin{eqnarray}
  m(R_f)&=& 2\pi\rho_c R_0\int_0^{R_f/R_0}\frac{x}{1+x^2}dx\label{e:mRf}\\
        &=&
  \begin{cases}
    \frac{2m_{cr}}{1-p/2}\left(\left(\frac{\rho_c}{\rho_{ext}}\right)^{(2-p)/p}-1\right)&\mbox{ for }p\neq 2\\
    2m_{cr}\,\ln\left(\left(\frac{\rho_c}{\rho_{ext}}\right)^{2/p}\right)&\mbox{ for }p=2\nonumber
  \end{cases}
\end{eqnarray}
Note that $m(R_f)>0$ for all values of $p > 0$, but that $m$ will not converge for $p<2$. Yet, this is of minor concern here,
since the radius $R_f$ will always be finite by construction (see eq.~\ref{e:Rf}). For $p=4$, $m(R_f\rightarrow\infty)\rightarrow m_{cr}$.
The initial condition for equation~\ref{e:dmdt} is set by assuming an initial central
density $\rho_c$, and a (constant) ambient density. 
The central density as a function of $m$ is given by
\begin{equation}
  \rho_c = 
  \begin{cases}
    \rho_{ext}\left(1+\left(1-\frac{p}{2}\right)\frac{f}{2}\right)^{1/(2-p)}&\mbox{ for }p\neq 2\\
    \rho_{ext}\exp\left(\frac{f}{2}\right)&\mbox{ for }p=2,
  \end{cases}
\end{equation}
where we used equation~\ref{e:fcrit}.

\begin{figure}
  \includegraphics[width=\columnwidth]{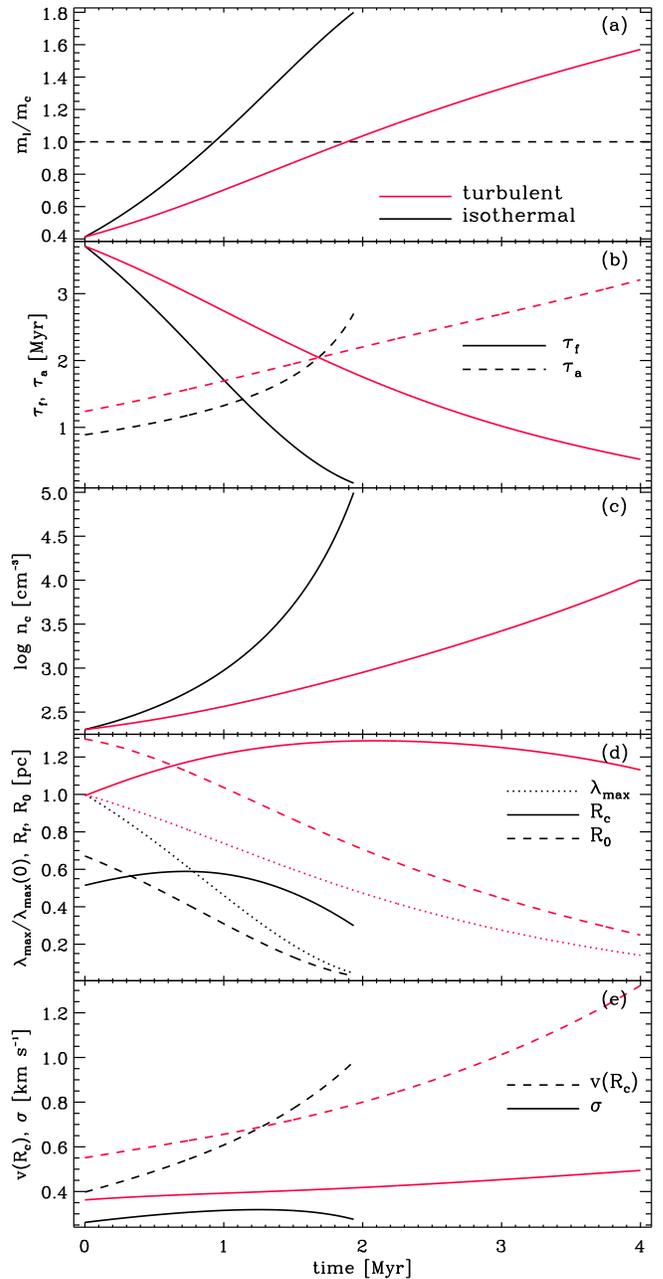}
  \caption{\label{f:macc_isotrb}Filament properties against time for the isothermal (black) and the turbulent (red) case.
           (a) Line mass over critical value (eq.~\ref{e:mlinecrit}). (b) Fragmentation
           timescale $\tau_f$ (solid line) and accretion timescale $\tau_a$ (dashed line).
           (c) Central filament density $n_c$. (d) The filament radius $R_f$ (eq.~\ref{e:Rf}) and
           the core radius $R_0$ (eq.~\ref{e:R0}) in parsec, and the wavelength of the most
           unstable mode (eq.~\ref{e:lmax}) normalized to its initial value of $4.5$~pc. (e) accretion velocity $v(R_f)$ (dashed line) and turbulent
           velocity $\sigma$ driven by accretion (solid line; see eq.~\ref{e:sigma} and \S\ref{ss:accturb}).}
\end{figure}

Figure~\ref{f:macc_isotrb} (black lines) summarizes the results of the integration for $p=3$
\citep[i.e. flatter than isothermal, and steeper than values observed by ][ see, however, \citealp{2011A&A...533A..34H} regarding 
evidence for isothermal profiles]{2011A&A...529L...6A},
$R_{ref}=2$pc, $n_0=100$~cm$^{-3}$, an initial central density of $n_c(0)=200$~cm$^{-3}$, and for $T=10$~K. 
For this parameter choice, the filament reaches criticality after $\approx 1$~Myr (Fig.~\ref{f:macc_isotrb}a).
Accretion dominates the filament evolution until shortly after that time (Fig.~\ref{f:macc_isotrb}b, black lines).
Other choices of $p$ give similar results.

The core radius $R_0$ (Fig.~\ref{f:macc_isotrb}d, dashed lines) decreases monotonically, 
because of the growing central density (Fig.~\ref{f:macc_isotrb}c). The filament radius $R_f$ (defined as the radius at which
the density profile drops to the background density) increases to a maximum, and then drops, since for large $\rho_c$, 
$R_f\propto \rho_c^{-1/6}$, and for small $\rho_c>\rho_{ext}$, $R_f\propto \rho_c^{1/2}$.

Also shown in Figure~\ref{f:macc_isotrb}d is the most unstable wave mode $\lambda_{max}$, normalized to its initial value of $\approx 4$~pc.
With increasing central density, $\lambda_{max}$ decreases.

\subsection{Effects of Accretion-driven Turbulence}\label{ss:accturb}
\citet{2010A&A...520A..17K} argue that in many astrophysical objects turbulence can be driven
by accretion. For molecular clouds, this point has been made by various groups based on simulations
of flow-driven cloud formation 
\citep[][see also \citet{2008MNRAS.385..181F} for a more systematic approach]{2007ApJ...657..870V,2008ApJ...674..316H}. 
In this scenario, turbulence is a consequence of the formation process 
initially, while at later stages, global gravitational accelerations drive ``turbulent'' motions.

\citet{2010A&A...520A..17K} estimate the level of turbulence driven by accretion (their eqs. 2, 3, and 23).
For the purposes here, the characteristic length scale is $2R_f$, the 
accretion velocity $v(R_f)$, and the driving efficiency $\epsilon=0.1$ (see their eq. (23)). 
Then, the turbulent velocity dispersion is given by
\begin{equation}
  \sigma=\left(2\epsilon R_f v^2(R_f)\f{dm/dt}{m(t)}\right)^{1/3}.
  \label{e:sigma}
\end{equation}
The choice of $\epsilon=0.1$ errs on the generous side -- \citet{2010A&A...520A..17K} quote
values of a few percent.

Figure~\ref{f:macc_isotrb}e (black lines) shows the velocity dispersion $\sigma$ against time 
(solid line), and the accretion
velocity at $R_f$ for the filament discussed above (dashed line). The velocity dispersion $\sigma$ 
is a factor of a few larger than the sound speed at $T=10$K used to evaluate $R_0$.
For $\epsilon=0.05$, $\sigma$ is on the order of the sound speed.

Yet, the above estimate of internal motions driven by accretion does not account for the effect
of an increased $\sigma$ on the critical line mass $m_{cr}$ and on the core radius $R_0$. 
The latter will affect the line mass $m$ of the filament, and thus
the accretion rate. Since the internal motions are noticeably driven by infall at a factor
of a few, a closer look at this backreaction seems in place. That the above discussion is 
inconsistent because of the missing effect of $\sigma$ on the filament growth is also demonstrated
by the shape of $\sigma(t)$ (solid black line in Fig.~\ref{f:macc_isotrb}e): just from virial
considerations, $\sigma$ should increase with $m$.  

Replacing the sound speed used to calculate $R_0$ by the velocity 
dispersion $\sigma$ has several consequences\footnote{This is often referred to as ``turbulent support''. As has been pointed 
out repeatedly, supersonic isothermal turbulence {\em cannot} support a gravitationally unstable region already for the simple 
reason that it leads to local fragmentation first \citep[][although \citet{2005MNRAS.361....2C} point out that the earlier
picture of a local reduction of the Jeans mass may be incorrect]{2000ApJ...535..887K,2001ApJ...547..280H}. A perfect example of this 
effect can be seen in the simulations of turbulent, isolated clouds by \citet{2005MNRAS.359..809C}. What is commonly 
referred to as ``support'' is in fact the effect of redistributing the densities in a region, leading to low volume filling factors and
-- eventually -- thus to lowered star formation efficiencies.}.
Since $R_0$ is needed to determine not only the critical line mass
but also the line mass of the accreting filament (eq.~\ref{e:mRf}), the accretion rate (eq.~\ref{e:dmdt})
will now depend on $\sigma$, and thus on itself. A root finder can be used to determine $m(\sigma)$, and
thus the accretion rate $\dot{m}$.

As the red lines in Figure~\ref{f:macc_isotrb}a show, the ratio $m/m_{cr}$ increases now at approximately half
the rate, while the central density grows even more slowly. After $4$~Myr, the filament has reached
$n_c = 10^4$~cm$^{-3}$, as compared to $\sim 1.8$~Myr for the isothermal case. Also note that the isothermal
$n_c$ begins to diverge, while the turbulent $n_c$ does not. This is a direct consequence of continued
accretion, driving internal motions and fragmenting the filament, as can be seen from 
Figure~\ref{f:macc_isotrb}e: $\sigma$ keeps rising for the turbulent case (red lines), driven by 
continued infall due to the ever-deepening gravitational potential. 

The fragmentation timescale $\tau_f$ (Fig.~\ref{f:macc_isotrb}b) and the most unstable wavelength
(Fig.~\ref{f:macc_isotrb}d) depend on $\sigma$ only through the central
filament density $n_c$. For the fragmentation timescale, this is obvious from equation~\ref{e:tau_f}, 
and for the most unstable mode (eq.~\ref{e:lmax}), we use the isothermal sound speed, not the velocity dispersion $\sigma$,
consistent with our interpretation of turbulence not being able to provide pressure support. 
The filament and the core radius $R_f$ and $R_0$ are roughly a factor of $2$ larger compared to the isothermal case. 
The filament radius does not
change significantly, while the core radius drops by a factor of $\approx 5$ to a few tenths of a parsec.

Summarizing, accretion-driven turbulence reduces the growth rates in all 
quantities, but it does not eventually stabilize the filament: the fragmentation timescale still wins
over the accretion timescale. Replacing the sound speed by the velocity dispersion $\sigma$ in $\lambda_{max}$
would increase the fragmentation scale at a given time, but this seems an improper thing to do.

\subsection{Accretion of Magnetic Fields}
\citet{2000MNRAS.311...85F,2000MNRAS.311..105F} discussed in great detail equilibrium configurations 
of cylinders with a variety of magnetic field geometries. Here, we are interested less in {\em stable} 
configurations, but to what extent magnetic fields affect the {\em growth} of the filament. 

The effect of magnetic fields can be included approximately by modifying the sound speed 
\begin{equation}
  \bar{c}_s^2\equiv c_s^2\left(1+\f{2}{\beta_0}\left(\f{n_c}{n_{c0}}\right)^{2s-1}\right),
  \label{e:magsound}
\end{equation}
with the initial plasma parameter 
\begin{equation}
  \beta_0 \equiv\f{2c_s^2}{c_A^2}=\f{8\pi c_s^2\rho_{c0}}{B_0^2}.
\end{equation}
Because of flux-freezing, a scaling of the magnetic field strength with
density is assumed, 
\begin{equation}
  B\propto n^s,
\end{equation}
which will depend on the field geometry through the exponent $s$. 
For fields along the axis of the filament
and for toroidal fields, $s=1$, with
\begin{equation}
  \sigma^2 = c_s^2\left(1+\f{2}{\beta_0}\f{n_c}{n_{c0}}\right).
  \label{e:magsound1/2}
\end{equation}
For a uniform field perpendicular to the filament axis,
$s=1/2$ under mass and flux conservation, resulting in
\begin{equation}
  \sigma^2 = c_s^2\left(1+\f{2}{\beta_0}\right).
  \label{e:magsound1}
\end{equation}
In this case, the magnetic pressure will stay constant, at its initial level.
Both cases $s=1/2,1$ will be explored.
As in the turbulent case, eq.~\ref{e:magsound} can be solved numerically, since the RHS depends
on $\bar{c}_s$ via the central density $n_c$.

\begin{figure}
  \includegraphics[width=\columnwidth]{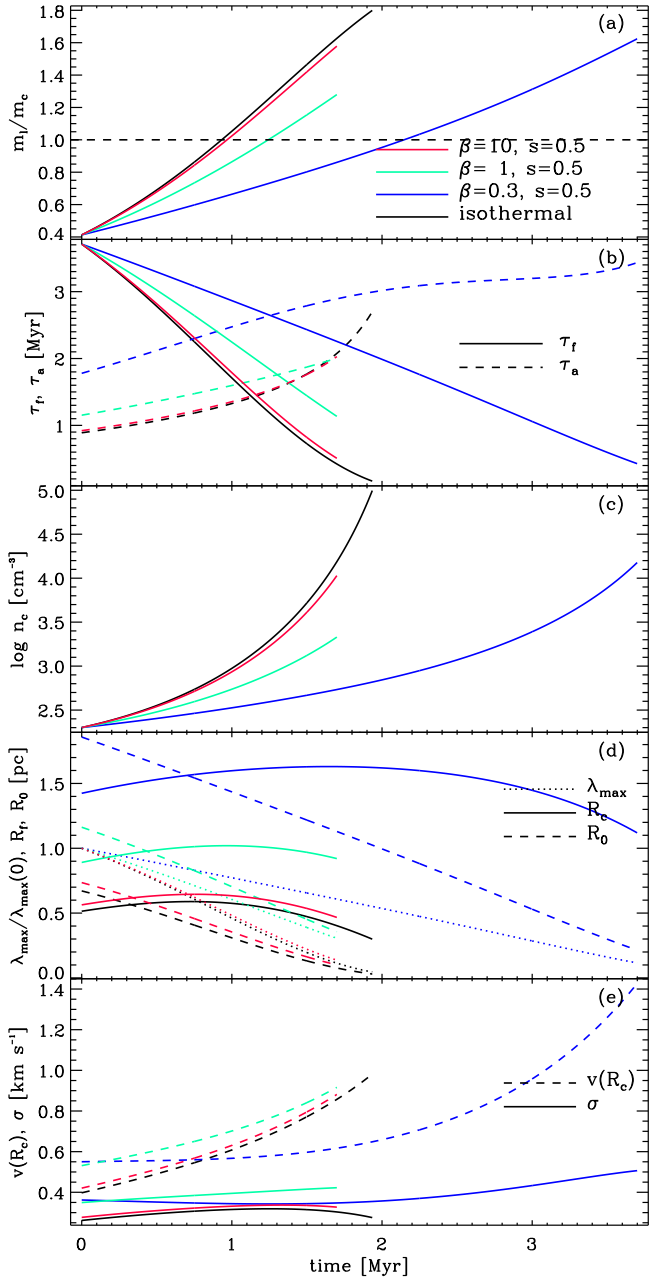}
  \caption{\label{f:macc_isomags05}Filament properties against time for the isothermal case (black), and the magnetic cases at $\beta_0=10$
           (red), $\beta_0=1$ (green), and $\beta_0=0.3$ (blue), 
           for a field perpendicular to the filament ($s=1/2$). 
           (a) Line mass over critical value (eq.~\ref{e:mlinecrit}). (b) Fragmentation
           timescale $\tau_f$ (solid line) and accretion timescale $\tau_a$ (dashed line).
           (c) Central filament density $n_c$. (d) The filament radius $R_f$ (eq.~\ref{e:Rf}) and
           the core radius $R_0$ (eq.~\ref{e:R0}) in parsec, and the wavelength of the most
           unstable mode (eq.~\ref{e:lmax}) normalized to its initial value. (e) accretion velocity $v(R_f)$ (dashed line) and turbulent
           velocity $\sigma$ driven by accretion (solid line; see eq.~\ref{e:sigma} and \S\ref{ss:accturb}).}
\end{figure}

\begin{figure}
  \includegraphics[width=\columnwidth]{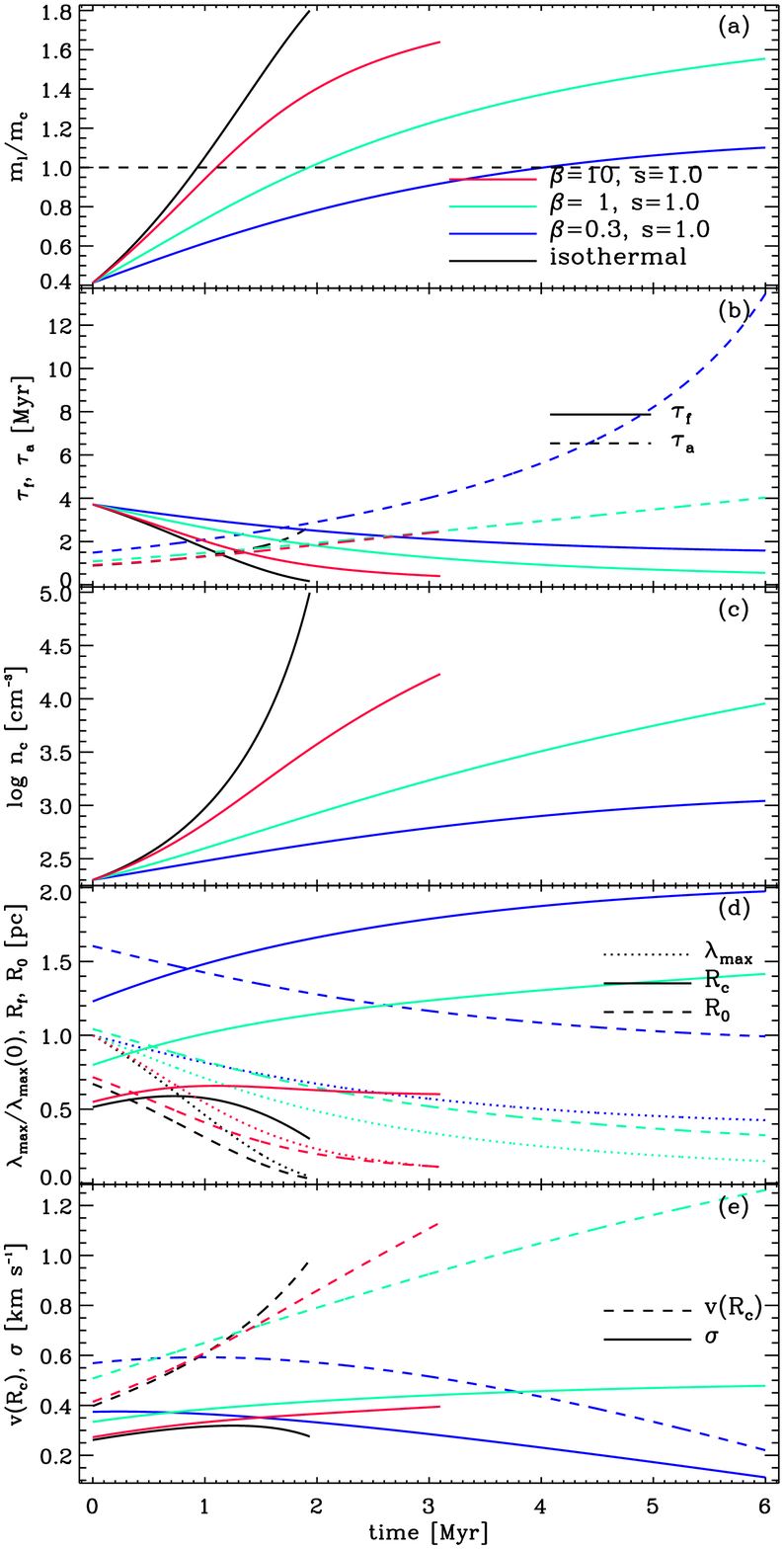}
  \caption{\label{f:macc_isomags10}Filament properties against time for the isothermal case (black), and the magnetic cases at $\beta_0=10$
           (red), $\beta_0=1$ (green), and $\beta_0=0.3$ (blue), 
           for a field parallel to the filament ($s=1$). 
           (a) Line mass over critical value (eq.~\ref{e:mlinecrit}). (b) Fragmentation
           timescale $\tau_f$ (solid line) and accretion timescale $\tau_a$ (dashed line).
           (c) Central filament density $n_c$. (d) The filament radius $R_f$ (eq.~\ref{e:Rf}) and
           the core radius $R_0$ (eq.~\ref{e:R0}) in parsec, and the wavelength of the most
           unstable mode (eq.~\ref{e:lmax}) normalized to its initial value. (e) accretion velocity $v(R_f)$ (dashed line) and turbulent
           velocity $\sigma$ driven by accretion (solid line; see eq.~\ref{e:sigma} and \S\ref{ss:accturb}).}
\end{figure}

The weak scaling ($s=1/2$, Fig.~\ref{f:macc_isomags05}) leads to qualitatively similar results as the isothermal case -- the magnetic curves have the
same shape as the isothermal ones. The only effect of the magnetic field in this case 
is to slow down the accretion, similar to the turbulent case. Fields parallel ($s=1$, Fig.~\ref{f:macc_isomags10}) to the filament, however, 
affect the accretion properties qualitatively. Both weak and strong 
fields ($\beta_0=10,1,0.3$) can slow down accretion, indicated by the flattening density and line mass curves at late times. 
The fragmentation timescale depends on the field strength through the central 
density $n_c$. The length scale of maximum growth (eq.~\ref{e:lmax}) depends
directly on the modified sound speed (eq.~\ref{e:magsound}) -- consistent with the notion that sufficiently strong magnetic fields 
can suppress gravitational
fragmentation\footnote{A different question is whether such strong fields can be reached in gravitationally contracting regions.}. 

While magnetic fields can affect the mass accretion onto the filament, they do not prevent it (at least not for
reasonable choices of magnetization). Fragmentation wins over accretion once the filament becomes critical. 
We forego the discussion of turbulence in combination with magnetic fields. There is numerical and analytical evidence
that turbulence combined with magnetic flux loss mechanisms (ambipolar drift and reconnection) efficiently
reduce the dynamical importance of magnetic fields 
\citep{1999ApJ...517..700L,2010ApJ...714..442S,2002ApJ...578L.113K,2002ApJ...567..962Z,2004ApJ...603..165H}.

\section{Effect of tidal forces on filament accretion}\label{s:tidal}
\citet{2009ApJ...700.1609M,2011ApJ...735...82M} discussed a detailed model of star formation regions of 
the hub-filament morphology (see a list of examples in \citealp{2009ApJ...700.1609M}, \S2).
He then explored
a model in which clumpy molecular cloud gas is compressed 
into a vertically self-gravitating  and density-modulated layer, 
preserving filamentary structure in equilibrium and in collapse. 
\citet{2010A&A...520A..49S} observed "hub-type" filaments around the massive star-forming filament DR21, including
infall/collapse signatures onto the massive cores. \citet{2012A&A...540L..11S} concluded from their analysis
of the Rosette filaments that clusters form at the position of filament mergers or crossings, consistent with
numerical simulations \citep{2004ApJ...605..800L,2007ApJ...657..870V,2013arXiv1304.1367C}.

As an alternative interpretation to the equilibrium models of \citet{2009ApJ...700.1609M,2011ApJ...735...82M} on the one hand 
and as an extreme interpretation of the turbulent view \citep{2012A&A...540L..11S} on the other
we assume that the gas around the massive filament is 
in free-fall, since the massive star formation region at the center  must dominate the gravitational energy budget of the system. 
The "fanning-out" filaments (or "gravitational streamers") may be formed by the stretching of over-dense gas parcels 
(``clumpy'' cloud structure) due to the tidal forces of the central filament. This interpretation is largely motivated
by the notion of ``focal'' points in finite, non-spherical, self-gravitating structures \citep{2004ApJ...616..288B}, 
i.e. by the fact that gravitational accelerations peak at the ends of filaments or ellipses. These locally strongly
varying accelerations not only lead to {\em local} compressions due to {\em global} gravity \citep{1983A&A...119..109B,2001ApJ...556..813L}, but they
also affect the ratio of the local to global collapse time, rendering structures of lower dimensionality more prone to suffer
local collapse before they collapse globally \citep{2011ApJ...740...88P}. As a corollary, they increase the global
free-fall time of the structure \citep{2012ApJ...744..190T,2012ApJ...756..145P}.

\begin{figure}
  \includegraphics[width=\columnwidth]{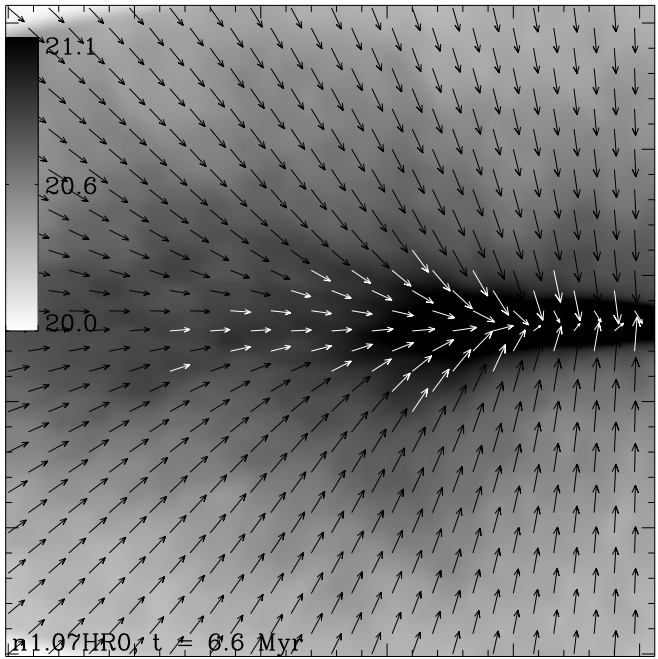}
  \caption{\label{f:focalvelo}Column density projection of 3D isothermal collapse of elliptical molecular cloud.
           Arrows indicate velocity vectors.}
\end{figure}

Figure~\ref{f:focalvelo} shows a column density projection (approximately $2$~pc across) of the 
three-dimensional isothermal collapse of an elliptical 
cloud. At this stage, the cloud has collapsed to a filament of $M=10^4$M$_\odot$, but ambient gas is still
being accreted. Specifically, at the end points of the filament (just one side is shown in Fig.~\ref{f:focalvelo}),
material is being accelerated parallel and perpendicular to the filament axis, leading to the ``velocity fan'' indicated
by the vectors. The initial density structure of the ambient gas is slighly clumpy, not filamentary, yet, the resulting
structures fanning out from the filament's end point are reminiscent of the hub filaments seen in observations.

\subsection{The Role of Tidal Forces}
The tidal accelerations on a gas parcel in free-fall towards a mass $M$ at a given distance $R$ are set by
\begin{eqnarray}
  a_T&=&\f{2GM}{r^2}\f{\Delta r}{r}\\
     &=& 87 \f{\Delta r}{r}\left(\f{M}{10^4\mbox{ M}_\odot}\right)\left(\f{r}{\mbox{pc}}\right)^{-2}\mbox{ km s}^{-1}\mbox{ Myr}^{-1},\nonumber
  \label{e:acctidal}
\end{eqnarray}
with the size of the parcel $\Delta r$, and its distance to the mass, $r$. For a distance $r=1$~pc, a size of $\Delta r=0.1$~pc,
and for a mass of $M=10^4$M$_\odot$, this results in a tidal acceleration of $0.87$km~s$^{-1}$ per $10^5$ years. 

\begin{figure}
  \includegraphics[width=\columnwidth]{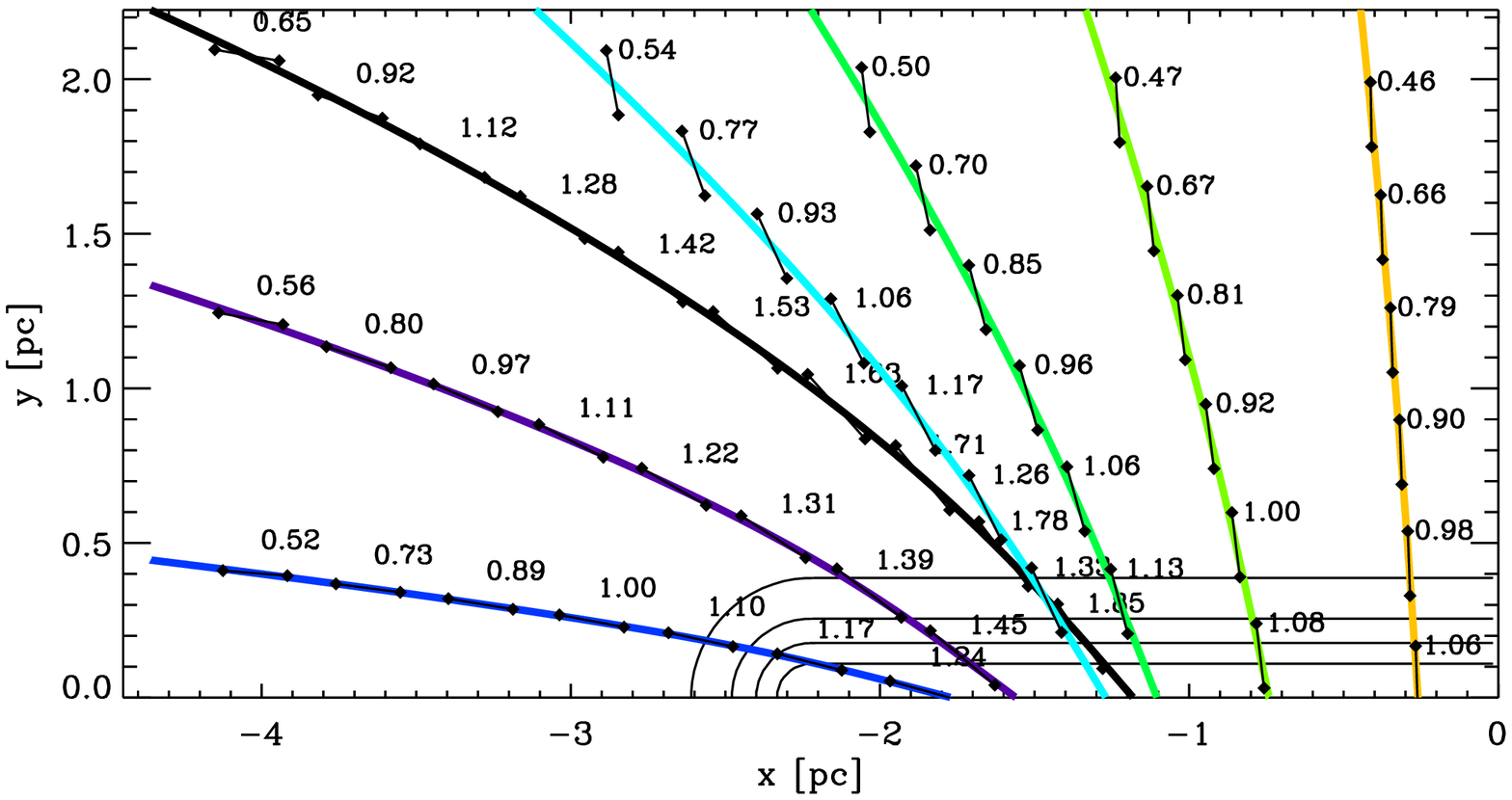}
  \caption{\label{f:ballistics}Trajectories for a fluid parcel under free-fall onto a filament, for $n_c=10^5$~cm$^{-3}$. 
           The filament -- modeled as
           a cylinder, see eq.~\ref{e:modelfil} -- is indicated by contour lines between $-2$ and $0$pc (only one quarter of
           the domain is shown). 
           Black line segments show the orientation of the fluid parcel's long axis. Numbers give times along the trajectory
           in Myr. Each of the fluid parcels starts at rest (see also Fig.~\ref{f:densitycomp}b). Line segments parallel with
           the trajectories are hard to discern.}   
\end{figure}

The effect of tidal forces on spherical fluid parcels can be estimated via a ballistic integration. The parcels are
initially at rest, accelerating towards the filament. The filament is modeled as an 
isothermal cylinder of total mass $M$, over a given length scale. 
The density profile is set by 
\begin{equation}
  n(x,y,z) = n_0+\f{n_c-n_0}{\left(1+\left(\f{R}{R_0}\right)^2\right)^{p/2}}
  \label{e:modelfil}
\end{equation}
with 
\begin{equation}
  R^2=
\begin{cases}
  x^2+y^2+z^2\mbox{ for } |x| \geq x_f\\
  y^2+z^2\mbox{ for } |x|<x_f.
\end{cases}
\label{e:modelfilr} 
\end{equation}
Here, $x_f=2$pc is the filament's half length, i.e. the filament extends from $-x_f$ to $x_f$. The background density $n_0=10^2$~cm$^{-3}$,
and central filament densities of $n_c=10^4,10^5$~cm$^{-3}$ are explored. The core radius is set to $R_0=0.1$~pc, and $p=3$.
Only one quadrant of the filament is shown in Figure~\ref{f:ballistics}, but the forces are calculated 
using the whole filament. The fluid parcels themselves are pressure-less, thus they can suffer infinite compression. 

We note the following from Figure~\ref{f:ballistics}: First, all parcels are being stretched along their trajectory, as 
indicated by the line segments giving the orientation of the major axis of a parcel. Second, the parcels hit the filament 
inwards of its endpoint (as a reminder, the center of mass resides at $x=0,y=0$). Third, with the stretching comes an 
overall acceleration of the parcel. And finally, all trajectories are curved towards the filament.

\subsection{Predictions}

\subsubsection{Curvature of Trajectories}
The most obvious difference between Figure~\ref{f:ballistics} on the one hand, and observations of gravitational streamers (or hub-filaments) 
as well as gas-dynamical models on the other (Fig.~\ref{f:focalvelo})
is that the ballistic trajectories curve {\em towards} 
the filament, while the observed and gas-dynamical ones tend to merge with the filament. 
A closer look at Figure~\ref{f:ballistics} provides the following information: Sampling the $(x,y)$-plane 
with more trajectories shows that the trajectories in the vicinity of the black line actually cross. Though the fluid parcels
are treated as pressure-less here, pressure forces would resist this compression perpendicular to the trajectory, 
preventing the crossing. Thus, a parcel's motion towards $y=0$ would be decelerated, leading to a reduced velocity 
component in $y$, and hence to an effective curving of the  trajectory such that it merges with the filament.

\subsubsection{Gas Density along Trajectory}\label{sss:gasdensity}
\begin{figure}
  \includegraphics[width=\columnwidth]{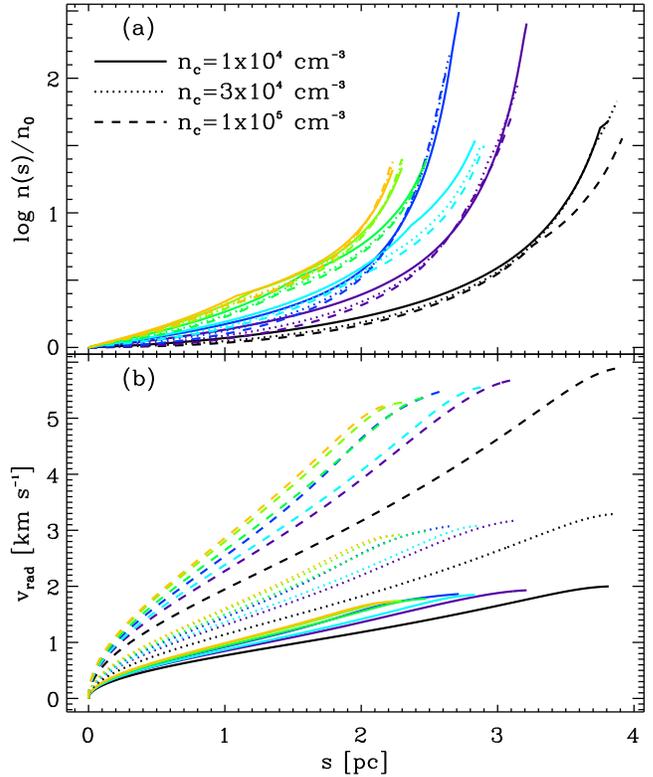}
  \caption{\label{f:densitycomp}Gas evolution along ballistic trajectories, for the three central filament
           densities $n_c=10^4,3\times10^4,10^5$~cm$^{-3}$ as indicated. (a) Density compression ratios
           along all trajectories of Fig.~\ref{f:ballistics}. 
           (b) Parcel velocities along all trajectories.} 
\end{figure}

Observed hub filaments are more pronounced closer the central structure 
\citep[see ][for a list of hub objects]{2009ApJ...700.1609M}.
Hence, the (column) density within the filaments is increasing. Thus, one would expect from the model
that the fluid parcels are not only be stretched, but also compressed. The inverse of the volume compression ratio 
$V(0)/V(t)$, which is equivalent to the density compression, for all trajectories is shown in Figure ~\ref{f:densitycomp}a. 
Compression is highest for trajectories with a strong component along the filament axis, while it is less pronounced 
for trajectories perpendicular to the filament. Trajectories along the filament axis (i.e trajectories that terminate at 
one of the filament's ends) effectively see accelerations $\propto R^{-2}$, while trajectories terminating further in are 
subject to accelerations $\propto R^{-1}$, with correspondingly weaker tidal forces.

\subsubsection{Line-of-Sight Velocity Signatures}
\begin{figure*}
  \hfill
  \includegraphics[width=\textwidth]{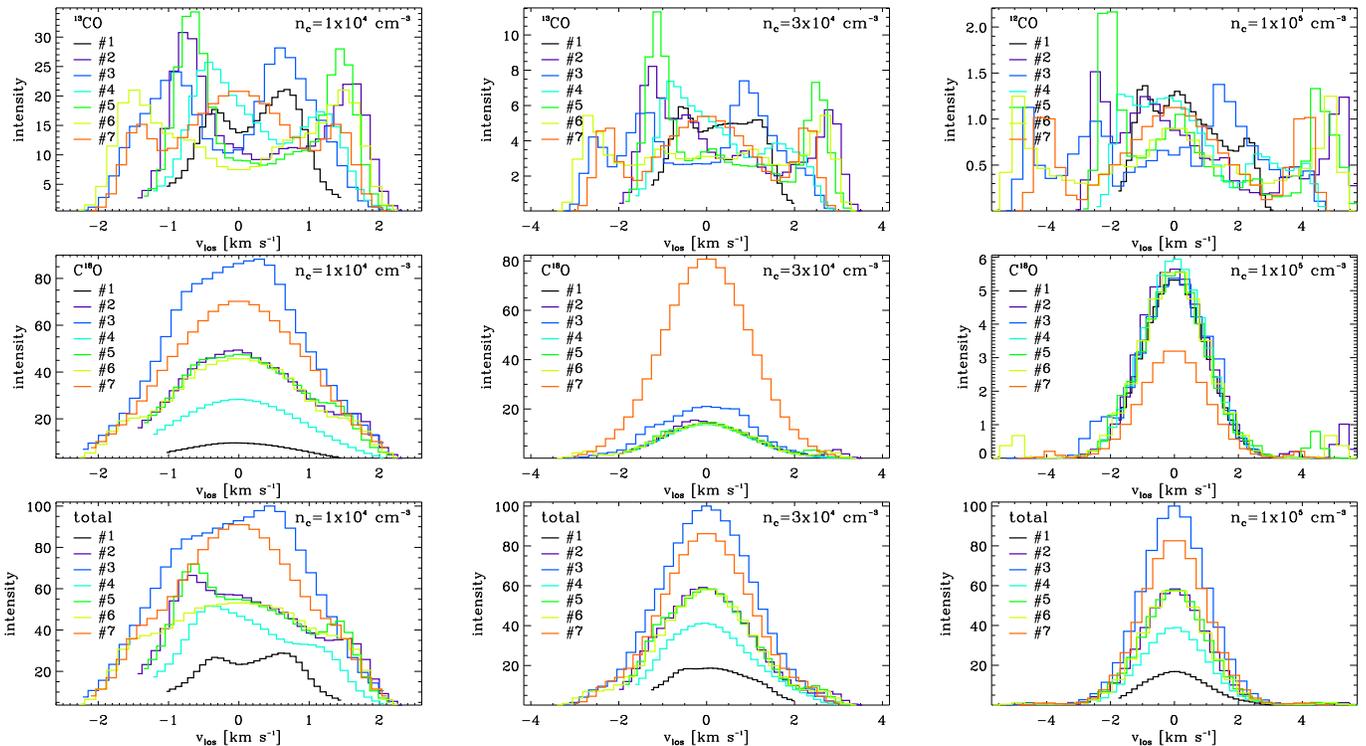}
  \caption{\label{f:vrad}Spectra for two density ranges (labeled $^{12}$CO and C$^{18}$O; see text),
           and for the full density range ("total"), calculated along lines-of-sight shown in Fig.~\ref{f:speclines}.
           Columns correspond to the central filament densities of $n_c=10^4,3\times10^4,10^5$~cm$^{-3}$ as indicated.
           The intensity units are arbitrary and are given in percent relative to the maximum intensity.}
\end{figure*}

\begin{figure}
  \includegraphics[width=\columnwidth]{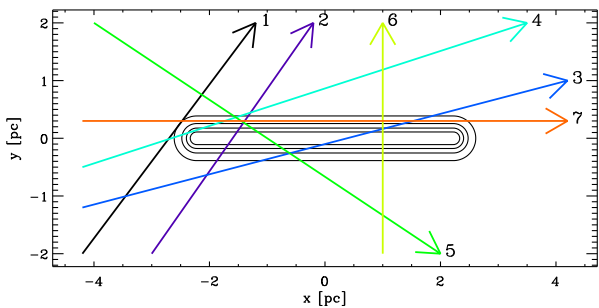}
  \caption{\label{f:speclines}Locations of lines-of-sight used for the spectra shown in Figure~\ref{f:vrad}.} 
\end{figure}

Figure~\ref{f:densitycomp}b shows the parcel velocities along all trajectories. These are consistent with the estimate in
eq.~\ref{e:acctidal}. Yet, an observer would not see the trajectories, but radial velocities along lines of sight
at a given time. Figure~\ref{f:vrad} provides an "observer's view" of the velocity structure around the filament. To derive
the spectra, lines-of-sight (as shown in Fig.~\ref{f:speclines}) are calculated through the filament and the ambient, infalling gas. Arrows 
in Fig.~\ref{f:speclines} indicate
the viewing direction. Spectra are determined assuming optically thin tracers corresponding to two density ranges 
($10^2<n<10^3$ and $10^3<n<10^4$), loosely identified as $^{12}$CO and C$^{18}$O. An initial parcel density
$n=10^2$~cm$^{-3}$ has been assumed. The sound speed in the accreting gas is set to
$c_s=0.2$~km~s$^{-1}$, and to $1$~km~s$^{-1}$ in the central filament, mimicking
the effect of accretion-driven turbulence. 
The emissivity is assumed to be proportional to the local density enhancement due to fluid parcel 
compression (see \S\ref{sss:gasdensity}). Emission from the infalling gas and from the central filament contribute to the spectra,
providing an estimate for the visibility of the predicted velocity structures given the presence of the main filament.

The bottom row of Figure~\ref{f:vrad} shows the spectra over the full density range, i.e. all material contributes to the emission.
Starting with the case $n_c=10^4$~cm$^{-3}$ (bottom left), we first note that the
lines-of-sight \#6 and \#7 are just for demonstration: the spectra are perfectly symmetrical around $v_{los}=0$ by construction. 
Line-of-sight \#5 starts out along the direction of maximum compression within the $R^{-2}$ regime 
(see discussion in \S\ref{sss:gasdensity}), intersects 
the filament, and then transverses a region of moderate acceleration (within the $R^{-1}$ regime). This is mirrored in the 
spectrum: the positive velocities peak around $v_{los}=1.5$~km~s$^{-1}$, while the negative velocities show a peak at 
$0.9$~km~s$^{-1}$. Line-of-sight \#4 sees a compression at lower negative velocities (corresponding to the far side), while
at positive velocities, it crosses a region of stronger compression and stronger acceleration.

Lines-of-sight \#1 and \#2 offer  comparative diagnostics: they run through the domain in parallel, with \#1 grazing the filament,
and \#2 intersecting. The spectrum along \#1 has a velocity range narrower by a factor of $2$ over spectrum \#2, and overall lower
emission. 
Line-of-sight \#2 is far enough in the filament that it picks up the emission from the filament itself, and that it 
mainly tracks the modest density compression gas
that falls onto the filament. Yet, given its orientation, the radial velocity forms a small angle with the infall velocity.
For \#1, because of the curvature of the trajectories, the angle between radial and infall velocity is larger, thus, the velocity
spread is smaller. For a more comprehensive picture, position-velocity plots will be useful (\S\ref{ss:pvmaps}).

With increasing central filament density (center and right column of Fig.~\ref{f:vrad}), 
the total emission is being dominated by the filament, and dynamical signatures are indiscernible
for $n_c=10^5$~cm$^{-3}$. Yet, the assumption of one tracer covering the full density range is unrealistic. The upper rows in
Figure~\ref{f:vrad} show the spectra for material emitting within specific density ranges.
For our parameter regime, the gas traced by the low-density tracer shows the strongest kinematic differences (top row). 
The central filament has been
nearly completely suppressed, leaving the infall signatures visible. With increasing filament densities, the infall peaks (which are located 
closest to the filament) appear at increasingly higher velocities. Higher-density tracers (second row) start to pick up the
central filament again.

\subsubsection{Centroid Velocity Signatures}\label{ss:pvmaps}
\begin{figure*}
  \includegraphics[width=\textwidth]{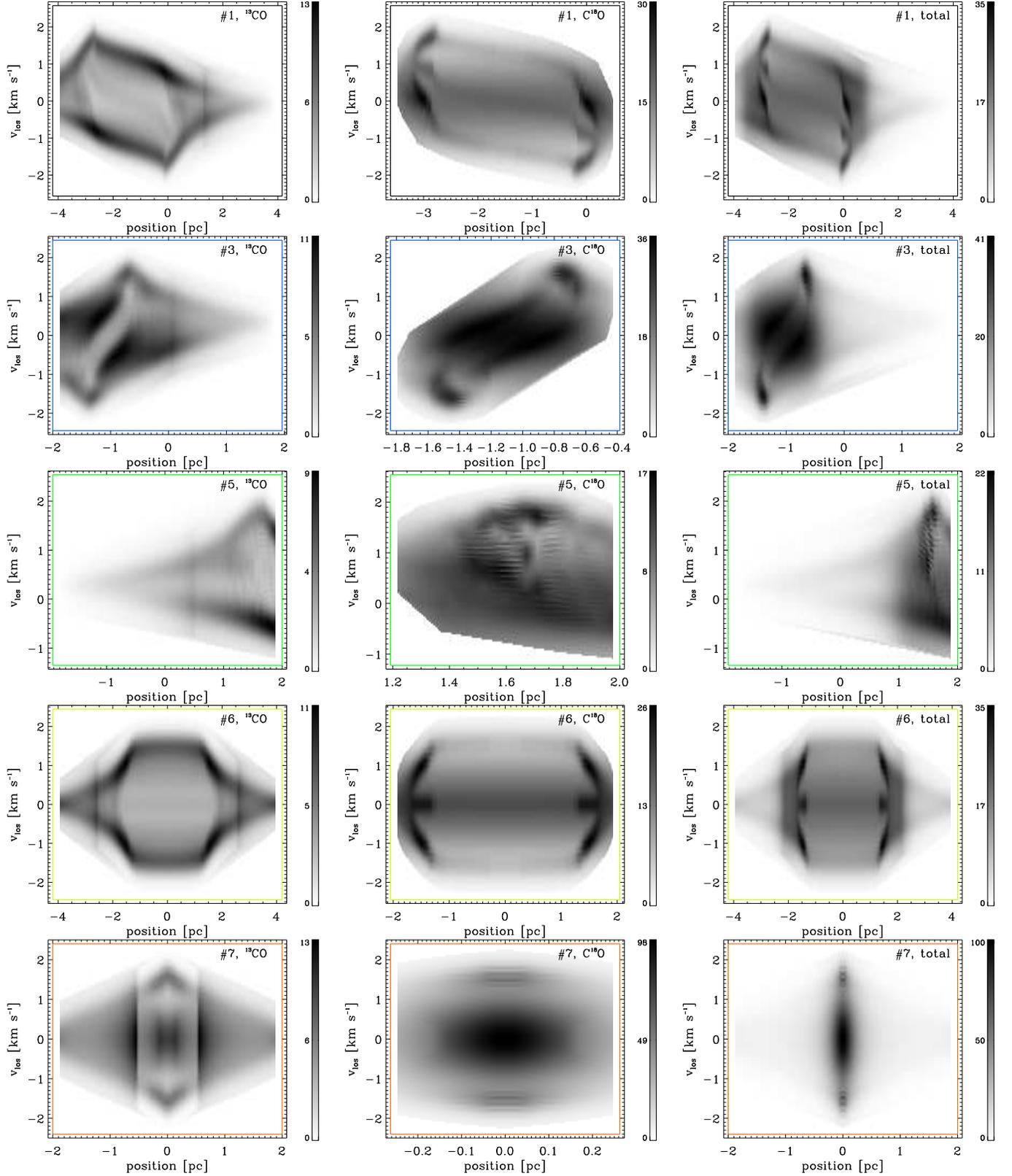}
  \caption{\label{f:posvela}Position-velocity plots for the lines-of-sight shown in Fig~\ref{f:speclines}, for $n_c=10^4$~cm$^{-3}$. 
           Rows contain the lines-of-sight as indicated (see Fig.~\ref{f:speclines}), and each column shows one tracer ($^{12}$CO, 
           C$^{18}$O, and total column density). Small-scale noise in the maps is due to unresolved velocity gradients (see text).
           Lines-of-sight \#1 and \#6 have been sampled along the $x$-direction (see Fig.~\ref{f:speclines}), and lines \#3, \#5 and \#7 along
           the $y$-direction. The intensity 
           units are arbitrary and are given in percent relative to the maximum intensity. Line-of-sight numbers and colors are given
           in each panel.}
\end{figure*}

Figure~\ref{f:posvela} provides a qualitative picture of the gas kinematics around a filament in free-fall accretion. 
Each panel corresponds to a position-velocity (PV) plot for lines-of-sight parallel to those indicated in Figure~\ref{f:speclines}.
The intensity units are arbitrary, and they result from weights proportional to the local density enhancements.
Lines-of-sight with slopes flatter than $1$ in Figure~\ref{f:speclines} have been sampled along the
$y$-axis, and those steeper along the $x$-axis, resulting in different ``position''-ranges in Figure~\ref{f:posvela}.
Also, position ranges are limited by the presence of a tracer, thus, spatial scales vary. 
Note that the plane of sky in these PV plots is (in Fig.~\ref{f:speclines}) perpendicular to the line-of-sight {\em and} the filament.
Small-scale noise in the maps is due to unresolved velocity gradients and can be ignored. Interpolating between cells with velocity
differences larger than the thermal width can remove these artifacts. Likewise, detailed structures within the PV envelope should
not be over-interpreted because of the very basic assumptions regarding the density-dependence of the tracers.

Lines \#1 and \#3 (Fig.~\ref{f:posvela}) show the density enhancements in the low-density tracers of the near and far infalling material, 
at relatively high velocities, i.e. close to the filament proper. At position$=-4$~pc (line \#1),   
material close to the left end of the filament is traced, while at position$=0$~pc, a contribution from near material due to
cylindrical infall (minor compression) and far material ($v_{los}<0$) towards the right end of the filament (thus compressed)
is being picked up. 
Line \#5 is an extreme case, showing the strong density enhancement of the near material, while the far side falls in at lower densities 
and velocities (especially in projection). The PV maps for lines-of-sight \#6 and \#7 show the symmetrical pattern to be expected 
for the infall velocities projected on the line-of-sight.

Higher-density tracers tend to map gas closer to the filament, and thus at higher velocities. For increasing central filament densities
(not shown), the filament itself becomes more prominent. 

\section{Discussion}\label{s:discussion}

\subsection{Gas Accretion onto Filaments}
\S\ref{s:accretion} demonstrates that a complete picture of filamentary molecular cloud evolution needs to
consider the environment of these clouds. Given their masses, they generate gravitational potential wells deep enough
that it seems unrealistic to neglect accretion. For typical cloud parameters, the cloud mass can double within a Myr, given
the availability of a mass reservoir. 


\subsubsection{Differences to Fischera \& Martin (2012)}\label{sss:differences}
In their analytical study of isothermal hydrostatic cylinders, \citet{2012A&A...542A..77F} develop a model of filamentary
molecular clouds pressurized by the ambient medium. 
They focus on the {\em structure} of the filament itself, depending on the external pressure. Instead of 
imposing a density profile by varying the exponent $p$, they argue that
reducing the external pressure will lead to a steepening of the profile (see their Fig.~3), from $p<4$ to $p=4$ for zero (vacuum) pressure.
Though the profiles indeed flatten with increasing external pressure, the value of $p$ -- and thus any power-law fit -- is actually unaffected
(Heitsch, in preparation). 

The very assumption that the filaments are embedded in a parental molecular cloud mandates not only that the density outside
the filament is non-zero, but also, that there is a (possibly substantial) external pressure component. \citet{2011A&A...530A..64K} discuss
evidence for dense structures in molecular clouds to be pressure-confined by the ambient cloud mass -- by themselves, the cores would be unbound. 
Yet, here, we are concerned with the {\em accretion} behavior of the filament, and less with the filament's radial structure, and thus leaving the
discussion of whether equilibrium considerations are appropriate for another time.
Instead of deriving the filament structure from first principles, we {\em assume} that we can subsume any physical processes leading to a flattening
of the profile into setting a profile index, consistent with observed values. Again, since the profile is only used to determine the line mass $m$,
this "morphological" approach seems acceptable -- our results regarding filament growth through accretion do not change 
qualitatively with different values of $p$. 
Flatter than isothermal ($p=4$) profiles have been ascribed to magnetic fields \citep{2000MNRAS.311...85F},
accretion, and non-isothermality \citep[][Heitsch, in preparation]{1999ApJ...515..239N}.
The advantage of imposing $p$ is that the model should be applicable to a variety of physical processes resulting in different values of $p$. Thus,
the model is to some extent agnostic on contentious issues like the admissability of hydrostatic equilibrium considerations in a turbulent environment.
Obviously, imposing $p$ is also a disadvantage -- the physical properties of the filament {\em together} with its environment are not fully explored. 

\subsubsection{Accretion and ``Nessie''}\label{sss:nessie}
\citet{2010ApJ...719L.185J} estimate typical core separations of $4.5$pc for the ``Nessie''-filament [$(l,b)=(338.4,-0.4)$] of
a total length (at a distance of $3.1$kpc) of $80$pc. Assuming an isothermal cylinder, they conclude that 
the core separation is by a factor of $\approx 5$ larger than the estimated central density of $10^4$~cm$^{-3}$ would
warrant. The problem is directly obvious from Figure~\ref{f:macc_isotrb}c and \ref{f:macc_isotrb}d. At a central
density of $n=10^4$~cm$^{-3}$, the core separation should be $\approx 1$~pc, but \citet{2010ApJ...719L.185J} observe
$4.5$pc, corresponding to a much lower central density of (in our model) $200$~cm$^{-3}$. 

\citet{2010ApJ...719L.185J} suggest additional support mechanisms, such as magnetic fields
\citep{2000MNRAS.311...85F,2000MNRAS.311..105F}, and/or turbulence. As can be seen from Figure~\ref{f:macc_isomags10}c and
\ref{f:macc_isomags10}d, only the strong magnetic field case ($\beta = 0.3$) has a chance of reaching an equilibrium
solution (with zero infall). Yet, the densities and the fragmentation length still fall short by a factor of $\approx 2$.

In the context of the discussion in \S\ref{ss:accturb}, "Nessie" would still be accreting gas, possibly driving 
turbulence via accretion \citep{2010A&A...520A..17K}. 
{\em Yet while accretion-driven turbulence can slow down accretion and thus can delay the onset of gravitational fragmentation and
collapse, it cannot prevent it (Fig.~\ref{f:macc_isotrb}).} Even when starting with a turbulent velocity dispersion \citep[at rather generous efficiencies
of $10$\%,
see discussion by][]{2010A&A...520A..17K}, the accretion velocity is always larger and grows faster than
the turbulent velocity dispersion. This short-coming of a turbulent support scenario is independent of the fact that the turbulent
velocity field itself -- in order to give rise to a supporting pressure -- would have to be uniquely configured, which
is extremely unlikely to occur \citep{2004ApJ...616..288B}. We conclude that while accretion will
drive turbulent motions in a filament, and while it possibly will lead to fragmentation of the filament, it cannot lead
to turbulence that in turn can support the filament.  

"Nessie" could have fragmented at an earlier stage during its formation, with the sub-filaments still accreting
gas and thus growing. To reach the characteristic fragmentation scale of $4.5$~pc, a central density of
$200$~cm$^{-3}$ would be needed, i.e. the fragmentation would have had to occur very early during the formation indeed. 

Finally, Herschel studies of filaments \citep[e.g.][]{2010A&A...518L.102A,2011A&A...529L...6A,2013A&A...550A..38P} 
suggest that with higher resolution and sensitivity available, more cores and
fragmentation centers seem to appear, possibly indicating that the $4.5$~pc quoted by  \citet{2010ApJ...719L.185J} may be an upper
limit for the fragmentation scale.

\subsubsection{Filament Evolution and FWHM($N_c$)-Correlations}
In their Herschel study of dust filaments in Aquila, Polaris and IC5146, \citet{2011A&A...529L...6A} conclude from 
a radial profile analysis that their identified filaments may share a similar characteristic width of $\sim 0.1$~pc.
They also point out that the filament width does not depend on the central column density $N_c$,
even for the gravitationally unstable filaments in their sample. They argue  
that a turbulent filament formation mechanism as discussed by \citet{2001ApJ...553..227P} may explain the similar widths for
gravitationally {\em stable} filaments, and that for gravitationally {\em unstable}
filaments it could be a consequence of continuing accretion, assuming a virialized filament. 

To test the latter statement, $2800$ filament accretion models (see \S\ref{s:accretion}) were run, varying four parameters:
the profile exponent $1.25<p\leq 3$ (see eq.~\ref{e:rhoR}), the driving efficiency $0.5\%\leq \epsilon \leq 5\%$ (see eq.~\ref{e:sigma}), 
the background density $10^2\leq n_0 \leq 3\times10^3$~cm$^{-3}$, and the ratio between the initial central density and the background density,
$1.1\leq n_c(0)/n_0\leq 5.0$. The choice of ranges is to some extent dictated by getting stable solutions.
The physics were restricted to accretion-driven turbulence, since the isothermal and the
magnetized cases do not leave any option to break the FWHM($N_c$) correlation. As a corrollary, it is unlikely that the infall is 
magnetically dominated, at least not at the simplistic level considered here \citep[also, see ][for a discussion of magnetically
dominated accretion in Taurus]{2013A&A...550A..38P}. 

\begin{figure*}
  \includegraphics[width=\textwidth]{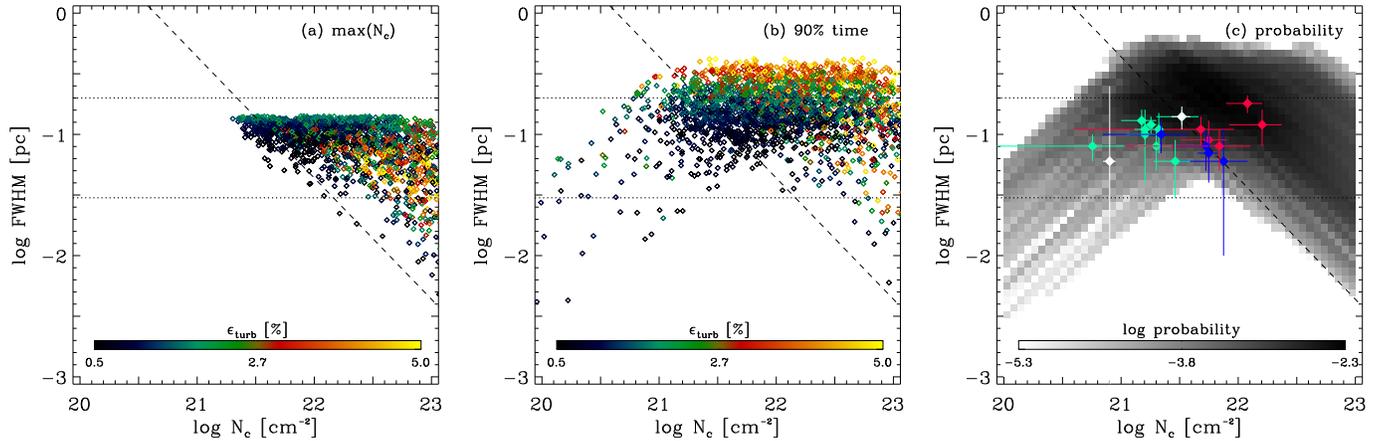}
  \caption{\label{f:arzoumanian}FWHM against column density for all model filaments. (a) Quantities determined at final stage (i.e.
           maximum column density), (b) time-sampled quantities, and (c) probability map.
           The plots can be directly compared to Figure~7 of
           \citet{2011A&A...529L...6A}. The range of observed FWsHM is given by the dotted lines, and the dashed line indicate the Jeans length,
           $\lambda_J=c_s^2/(G\mu m_H N_c)$. 
           The driving efficiency $\epsilon$ (see \S\ref{ss:accturb} and eq.~\ref{e:sigma}) is indicated by colors.
           In panel (c), observed filament values are plotted as diamonds, with colors indicating whether the filament contains YSOs (red), pre-stellar
           cores (blue), cores (green), or nothing (white). 
           Error bars indicate the uncertainties given by \citet{2011A&A...529L...6A}.}
\end{figure*}

Figure~\ref{f:arzoumanian} summarizes the test results. In panels (a) and (b), each data point represents the characteristic values of the 
FWHM and central column density for one model (thus, there are $2800$ data points). In Figure~\ref{f:arzoumanian}a, the values are picked
at the end of the model, i.e. for the final (maximum) column density. Thus, nearly all filaments are gravitationally unstable (towards large
$N_c$ with respect to the Jeans length, indicated by the dashed line), since
the integration proceeded beyond $m/m_{cr}=1$. For Figure~\ref{f:arzoumanian}b, the data points are randomly sampled, weighed by 
the time spent in a given column density bin. The positions in the $(\mbox{FWHM},N_c)$-plane are determined by the median of the column density
range within which the filament spends $90$\% of its time. Since the column density tends to grow more slowly at early stages, 
the early stages of the filament's growth are sampled. Thus, there is a good fraction of gravitationally stable filaments. Figure~\ref{f:arzoumanian}c
maps the probability to find a filament at a given position in $(\mbox{FWHM},N_c)$ space during its evolution.
The evolutionary tracks of single filaments can be guessed from the diagonal streaks towards the instability line, and parallel to
it for large column densities.

The results are analyzed with the expressions for the (column) density profile by \citet[][their eq.~1]{2011A&A...529L...6A}.
Since the exponent $p$ is one of the random parameters, we determine the constant $A_p$ for a range of exponents and fit it with a cubic. 
The FWHM is determined by
\begin{equation}
  \mbox{FWHM}=R_0\left(2^{2/(p-1)}-1\right)^{1/2}.
  \label{e:fwhm}
\end{equation}

If we choose to "observe" the model filament at its final stage (Fig.~\ref{f:arzoumanian}a), the upper envelope of the column 
density distribution is flat. There is a tendency towards lower FWHM with higher column densities, mirroring the expected effect
of filament contraction. Most of the filaments' FWsHM reside within the observed range (dotted lines). Nearly all data points
are located in the unstable regime -- only a small fraction of filaments never makes it to gravitational collapse. 

The time-sampled distribution (Fig.~\ref{f:arzoumanian}b) extends into the stable regime, and it suggests
consistency with the observed decorrelation between FWHM and $N_c$. Clearly, the initial (gravitationally stable) conditions of
the filaments are now contributing. The FWHM decreasing with smaller column densities is to a large extent an artifact of the
initial conditions: we choose a random combination of parameters and assume that the filaments have evolved profiles. In other words,
these models cannot address the question of filament {\em formation}. As in Figure~\ref{f:arzoumanian}a, there is a tendency towards
smaller FWHM at high column densities.

Finally, Figure~\ref{f:arzoumanian}c gives a summary view of the filament distribution and evolution. As for the time-sampled
case of panel (b), the filament trajectories in the stable regime (left of the dashed line) are possibly a result of our initial
conditions. Yet, over a large range of observed column densities, the FWHM stays flat. We overplotted the FWHM($N_c$) values
for selected filaments in IC~5146, drawn from \citet[][their Table~1]{2011A&A...529L...6A}. Colors indicate whether the filament
contains YSOs (red), pre-stellar cores (blue), cores (green), or nothing (white). 

We note several points: 

\paragraph{(a)} Overall, the models can reproduce a flat FWHM($N_c$) distribution. The flatness is a direct consequence of accretion-driven
turbulence. For low efficiencies $\epsilon$ (blue colors in Figs.~\ref{f:arzoumanian}a,b) a down-turn at high column densities is more 
readily observed. The isothermal case not including turbulence (not shown) results in a strong correlation between FWHM and $N_c$ similar
to that observed by \citet{2012A&A...542A..77F}.
The same effect can be seen for the magnetic cases (see Fig.~\ref{f:macc_isomags05}, \ref{f:macc_isomags10}): either, the scaling
of the field with density is too weak to affect the results (and thus yielding a strong correlation between FWHM and $N_c$), 
or the filament widths stay flat with $N_c$, but are an order of magnitude larger than observed.

\paragraph{(b)} At high column densities, the FWsHM decrease, as one would expect for gravitationally
contracting filaments. Yet this is inconsistent with Figure~7 of \citet{2011A&A...529L...6A}. At such high column densities, the filaments
are expected to fragment and thus form stars, resulting in structures that may not be recognized as filaments. Thus, if there were a down-turn,
it may be hard to observe. The following issue is also puzzling and may warrant further discussion: Ignoring the filaments with the two
highest column densities ($\log N_c=22.1,22.2$) results in a FWHM($N_c$) distribution that is clearly constrained by the Jeans length. In fact,
the filaments containing pre-stellar cores (blue) and YSOs (red) nearly all follow the Jeans length, while the remaining cores reside well to the left of the Jeans length.
As for a justification of ignoring those two data points, comparing Figure~3b and 4a of \citet{2011A&A...529L...6A} suggests that the filaments in 
question (\#6 and \#12) are in the region with highest confusion and thus present possibilities for overlaps. 
\citet{2012A&A...544A.141J} point out that overlaps and three-dimensional effects can affect the filament parameters derived from 
two-dimensional projections. As those authors show, what appears to be a continous filament in PPV space,
may well be a set of (isolated) structures in three dimensions \citep[see also Figs. 2, 3 and 4 of][]{2009ApJ...704.1735H}.

Obviously, in Figure~\ref{f:arzoumanian}c,
we have restricted ourselves to the readily available data set of IC~5146, leaving out the data for Aquila \citep{2010A&A...518L.106K}, 
on which Arzoumanians et al. conclusions largely rest.

\paragraph{(c)} The $(\mbox{FWHM},N_c)$-plane (Fig.~\ref{f:arzoumanian}c) is most densely populated at FWsHM somewhat larger than observed. The
"center of mass" of the model filaments resides more at $0.3$~pc, rather than at $0.1$~pc, as claimed by \citet{2011A&A...529L...6A}. One
possible reason could be the choice of the driving efficiency $\epsilon$: it is mostly the high efficiencies that contribute to the 
large FWsHM. Tests like these could be used to constrain observationally the turbulent efficiencies, but 
this would require a more detailed (and more realistic) approach to modeling the effect of accretion on the filament. 
The difference between modeled and observed FWsHM could also stem from 
how the FWsHM are measured. Finally, they could arise from observational selection effects. Though it is tempting to speculate whether observations 
tend to pick out filaments at their late stage of evolution (Fig.~\ref{f:arzoumanian}a) because of the highest available contrast,
or at earlier times because of more time being spent at lower column densities, the current study cannot give a definite answer. Yet, it 
seems likely that the observed filament samples are not unbiased. 

In summary, the filament models including accretion-driven turbulence can reproduce the observed decorrelation between filament
FWHM and column density. For higher column densities, a correlation eventually emerges, yet this stage would also be affected by
fragmentation and ensuing disruption. For low column densities, the formation mechanism may play a role (\S\ref{sss:initcond}).

\subsubsection{Validity of the Free-Fall Assumption}
The discussion hinges on the assumption that the gas is in free-fall towards the filament. This may -- at first sight --
seem an extreme assumption. Yet, for the sake of offering an alternative picture to the classical equilibrium
considerations, the free-fall assumption is meant as a counter-point. While there is tentative observational evidence
for accretion onto molecular clouds \citep[e.g. \citet{2009ApJ...705..144F}, see discussion in ][]{2010A&A...520A..17K}, the situation
is somewhat less clear in filaments, although recent work seems to support the claim
that there are global infall motions 
\citep[][see also Friesen et al., in preparation]{2005IAUS..227..151M,2010A&A...520A..49S,2012A&A...543L...3H,2013A&A...550A..38P}. 
Though somewhat smaller than predicted here, the observed kinematic signatures are on the order
of the virial velocities, which would make them mass-dependent. As long as the infall velocities depend on mass,
there is a basis to assume at least some free-fall component. 

\subsubsection{Initial Conditions}\label{sss:initcond}
The choice of the initial central filament density of $n_c(0) = 200$~cm$^{-3}$ is motivated by the goal to highlight
the evolution of the filament over a large dynamical range. As is clear from e.g. Figure~\ref{f:macc_isotrb}b, at such low densities,
the accretion timescales are comparable to turbulent or dynamical timescale, questioning whether considering the filament as unaffected
by its environment is realistic. Yet, choosing a higher $n_c(0)$ just results in a larger $f=m/m_{cr}$ as starting point.
As \citet{2012A&A...542A..77F} point out, $f$ can be used as a proxy for the time in the filament's evolution. 
Forming filaments of a few hundreds to thousands cm$^{-3}$ poses no problem in a turbulent
ISM, especially given (transient) converging flows and strong radiative losses 
\citep{2005A&A...433....1A,2006ApJ...648.1052H,2006ApJ...643..245V,2007ApJ...657..870V,2008ApJ...674..316H,2013arXiv1304.1367C},
even in the presence of magnetic fields \citep{2008ApJ...687..303I,2009ApJ...695..248H}, or, in a more realistic setting, by
wind-driven superbubbles \citep{2011ApJ...731...13N}.

\subsection{Effect of tidal forces on filament accretion}
Tidal forces around finite cylinders can lead to stretching and compression of infalling fluid parcels (\S\ref{s:tidal}). 
Interpreting the fluid parcels as ``clumpy'' molecular gas in the filament's vicinity, this effect could explain
the ``gravitational streamers'' or ``fans'' around star-forming filaments 
\citep[see][for a selection of objects, and also for an alternative explanation involving instabilites]{2009ApJ...700.1609M,2011ApJ...735...82M}. 

\subsubsection{Observational Tests}
The discussion in \S\ref{s:tidal}, specifically Figure~\ref{f:posvela}, could be used in two ways to 
interpret observations of a filament of known mass.
First, if one assumed that the gas around an observed filament is in approximate free-fall, then the variation
in the spectra with line-of-sight positions (i.e. the PV plots) could be used to determine the orientation of the 
filament with respect to the line-of-sight. Conversely, if the orientation were known, one could estimate to what 
extent the gas is actually in free-fall around the filament. At minimum, infall signatures could be searched for,
independent of whether this infall is in free-fall or not. Uncertainties are introduced by an assumed
starting position of the fluid parcels -- it is unlikely that they will be at rest. 

Figure~\ref{f:posvela} offer an observable test of the hypothesis that the ``fans'' are caused by tidal forces. Yet,
the analysis assumes  that the line of sight and the filament are located in the same plane, i.e. that a projection
of the line-of-sight has a parallel component to the filament. Thus, for a comparison of Figure~\ref{f:posvela} 
with observations, the observational spectra should be taken {\em along} the axis of the central filament (see Friesen et al., in preparation).
A more detailed study would include full three-dimensional effects, with line-of-sight components perpendicular
to the filament axis. This would also catch the signature of the cylindrical infall, if present. This complication
has been omitted for the sake of clarity, but it should be addressed in the future to enable a more robust
observational test.

\citet{2013ApJ...766..115K} find signatures of flows {\em along} filaments. Such dynamics internal to the central
filament have been neglected here, simplifying the interpretation of the PV plots. Since the infalling material
presumably shocks at high densities close to the filament, the internal motions are likely to be smaller than
the infall motions. 
Also not considered was the possibility that the tracers could very well be optically thick, although in view of
the large velocity gradients, this might be less of an issue.   

Given the free-fall velocities of a few km~s$^{-1}$ at the position of the filament, the gas could possibly shock. This
assumes a well-defined filament, and a more or less coherent infall. From the observed (column) density structures, it is
not clear how realistic such an assumption, is, but in any case, it could be tested with appropriate shock models 
\citep[e.g.][]{2012ApJ...748...25P}. 

\subsubsection{Validity of Ballistics Assumption}
The fluid parcels used to determine the amount of stretching and compression have been treated as pressure-less,
i.e. they can suffer infinite compression. Removing this simplification will lead to two effects. First, the 
density compression along a trajectory would be expected to be more gradual, instead of the strong compression
observed in Figure~\ref{f:ballistics}. Second, as the fluid parcels travel towards the central axis of the filament,
density and pressure will increase, thus decelerating the gas eventually. This would lead to a compression 
along a direction perpendicular to the filament axis, and thus to a flattening of the parcels, thus 
enhancing the appearance of fan-like structures.

\section{Conclusions}

In an attempt to elucidate the evolution of a filamentary molecular cloud including its environment,
the characteristic fragmentation and accretion timescales have been evaluated under the 
assumption of free-fall accretion. Though certainly an extreme view, it is supported by observational
evidence of virial-like infall of gas in massive star-formation regions. Moreover, the filamentary
structure of molecular clouds suggests pressure-less collapse to be dominant {\em at least during the
formation of a cloud}, a notion that in view of the thermal properties of the gas is not completely 
outrageous\footnote{From the observed internal dynamics of filaments it is obvious that they themselves are {\em not} in 
radial free-fall collapse.}.

The analysis presented in \S\ref{s:accretion} and Figures~\ref{f:macc_isotrb}-\ref{f:macc_isomags10} demonstrates
that the effects of accretion will play a role for any molecular cloud with physically reasonable
parameters. Magnetic fields and/or turbulence can slow down the accretion and thus delay the onset
of gravitational fragmentation, but they cannot prevent it. Including the effects of accretion-driven
turbulence slows down the build-up of the filament by approximately a factor of $2$. 

Observational evidence for a decorrelation between filament FWHM and peak column density \citep{2011A&A...529L...6A} can
be explained by continuing accretion -- as those authors had speculated.

Tidal forces around finite filaments can lead to stretching and compression of clumpy molecular 
gas being accreted onto the filament. The resulting gravitational streamers are reminiscent of ``fans'',
offering a possible explanation of such structures around star formation
regions. Position-velocity plots for such an accretion process offer the opportunity for a comparison with
observations of massive filaments, either determining the orientation of the filament with respect to the
line-of-sight (if free-fall is assumed), or -- if the orientation is known -- testing the hypothesis of
free-fall accretion. 

\acknowledgements The referee's report was {\em extremely} helpful -- thank you very much indeed to the anonymous
referee. I gratefully acknowledge support by NSF grant AST 0807305, and by NASA NHSC 1008. 
Insightful comments by and discussions with H.~Kirk, D.~Johnstone and L.~Hartmann 
are very much appreciated. I am indebted to S.E.~Ragan for explaining observational subtleties and
for a thorough reading of the manuscript.  

\bibliographystyle{apj}
\bibliography{./references}

\end{document}